\documentclass[letters,a4paper,fleqn,usenatbib]{mnras}

\usepackage[T1]{fontenc}
\usepackage{ae,aecompl}

\usepackage{graphicx}	
\usepackage{amsmath}	
\usepackage{amssymb}	

\newcommand{\eq}[1]{\begin{equation}  #1 \end{equation}}

\newcommand{\eqa}[1]{\begin{eqnarray}   #1 \end{eqnarray}}

\newcommand{\br}[1]{\left( #1 \right)}
\newcommand{\bc}[1]{\left\{ #1 \right\}}
\newcommand{\bb}[1]{\left[ #1 \right]}
\newcommand{\ba}[1]{\left\langle #1 \right\rangle}

\newcommand{\mx}[1]{\mathbfss{#1}}
\newcommand{\mxg}[1]{\mbox{\boldmath{$#1$}}}

\newcommand{\vek}[1]{\mbox{\boldmath $#1$}}

\title[Non-linear shrinkage covariance]{Non-linear shrinkage estimation of large-scale structure covariance}

\author[B. Joachimi]{Benjamin Joachimi\thanks{E-mail: b.joachimi@ucl.ac.uk}
\\
Department of Physics and Astronomy, University College London, Gower Street, London WC1E 6BT, UK
}

\date{Accepted XXX. Received YYY; in original form ZZZ}

\pubyear{2016}

\begin{document}
\label{firstpage}
\pagerange{\pageref{firstpage}--\pageref{lastpage}}
\maketitle

\begin{abstract}
In many astrophysical settings covariance matrices of large datasets have to be determined empirically from a finite number of mock realisations. The resulting noise degrades inference and precludes it completely if there are fewer realisations than data points. This work applies a recently proposed non-linear shrinkage estimator of covariance to a realistic example from large-scale structure cosmology. After optimising its performance for the usage in likelihood expressions, the shrinkage estimator yields subdominant bias and variance comparable to that of the standard estimator with a factor $\sim 50$ less realisations. This is achieved without any prior information on the properties of the data or the structure of the covariance matrix, at negligible computational cost.
\end{abstract}




\section{Introduction}

The covariance is an indispensable ingredient for inference from data, quantifying the varying levels of statistical uncertainty among the data points as well as their correlations. In many astrophysical situations the covariance is not known a priori and has to be determined from measurements along with the data, turning the elements of the covariance matrix themselves into random variables with associated errors. Cosmological data analysis faces a particular challenge in that only a single realisation of the data is available, and that treating representative subsamples of the data as quasi independent may be inaccurate due to long-range spatial correlations of the signals under investigation \citep{norberg09}. Therefore, one usually resorts to estimating a standard sample covariance matrix from simulated realisations of the data. 

The finite number of mock data realisations induces noise in the covariance estimate that propagates into the errors of inferred model parameters, which was first explicitly pointed out in a cosmological context by \citet{hartlap06} and subsequently investigated in detail \citep{taylor13,dodelson13,percival14,taylor14,sellentin16}. For Gaussian distributed data, the sample covariance follows a Wishart distribution. While e.g. the fields and derived statistics probed in cosmic large-scale structure (LSS) surveys follow strongly non-Gaussian distributions, the derived properties of the covariance inferred from the Wishart distribution turn out to still be applicable to very good approximation \citep{dodelson13,petri16}. A generic property of a Wishart matrix is that it becomes singular if the number of realisations used to estimate it, $N_S$, becomes less than the size of the data vector, $N_D$, prohibiting its use in likelihood analysis or least squares fitting, for which the inverse covariance is required. This implies that $N_S \gg N_D$ often computationally expensive simulations are required to determine the covariance, where $N_D \sim 1000$ will be readily surpassed by forthcoming cosmological surveys.

To lessen this computational bottleneck in the analysis, $N_D$ could be reduced via data compression, but good knowledge of the covariance is necessary to achieve near-optimal compression \citep[e.g.][]{tegmark97}. Innovative schemes to augment a given number of simulations via resampling techniques have been proposed as well \citep{schneiderm11b,escoffier16}. Alternatively, one can replace the sample covariance estimator with a generally biased one that has favourable noise properties. \citet{paz15} proposed to taper correlations far from the diagonal, which however requires a notion of distance between all elements of the data vector. \citet{padmanabhan16} investigated direct estimation of inverse covariance matrix elements, bypassing the sample covariance altogether. Linear shrinkage towards a modelled target \citep{pope08} or a constant correlation coefficient \citep{simpson16} has seen LSS applications. These estimators relied on the accuracy of the assumed model or structure of the covariance matrix, respectively, and were limited to a single global shrinkage intensity for all matrix elements, which can be suboptimal in improving the conditioning of the covariance \citep{ledoit12}. It is therefore timely to assess the performance of a non-linear generalisation of shrinkage covariance estimators, which has the added benefit of not relying on models or assumptions about the structure of the covariance matrix.

\section{Shrinkage estimator}

This work adopts the NERCOME\footnote{Non-parametric eigenvalue-regularised covariance matrix estimator} estimator recently proposed by \citet{lam16}, which in turn capitalised on earlier work by \citet{abadir14} and \citet{ledoit12}. Let $\mx{X}=(\vek{x}_1, \dots, \vek{x}_{N_S})$ be a $N_D \times N_S$ matrix of $N_S$ realisations (independent measurements) of the data vector, $\vek{x}_i$, each of length $N_D$. The data has covariance $\mxg{\Sigma}$, which however is unknown a priori. In the following it is assumed that the $\vek{x}_i$ are mean-subtracted and have potentially been normalised, i.e. their elements are given by $x_{\alpha i} = (x_{\alpha i}^{\rm raw} - \mu_\alpha)/n_\alpha$,\footnote{Latin indices denote different realisations while Greek indices cycle through the elements of the data vector.} where $\vek{\mu}$ is the vector of means (also estimated from the data) and $\vek{n}$ is a normalisation vector with noiseless entries. The standard sample covariance estimator reads
\eq{
\hat{\mx{{S}}} = \frac{1}{N_S-1}\; \mx{X}\; \mx{X}^\tau\;,
}
which is unbiased, $\ba{\hat{\mx{{S}}}} = \mxg{\Sigma}$. A key idea of NERCOME is to divide the dataset into two subsamples, $\mx{X} = (\mx{X}_1,\mx{X}_2)$, with $\mx{X}_1$ an $N_D \times s$ matrix and $\mx{X}_2$ an $N_D \times (N_S - s)$ matrix. The sample covariance can also be measured from each subset, denoted by $\hat{\mx{{S}}}_i$ with $i=1,2$. The estimator uses the diagonal decomposition of these estimates, $\hat{\mx{{S}}}_i =  \mx{U}_i\, \mx{D}_i\, \mx{U}_i^\tau$, where $\mx{U}$ is the matrix of eigenvectors and $\mx{D}$ is a diagonal matrix with entries $d_{\alpha\beta}=\delta_{\alpha\beta}\, \lambda_\alpha$, where the $\lambda_\alpha$ are the eigenvalues and $\delta$ is the Kronecker delta.

The NERCOME estimation process consists of three steps: 1. apply the basic estimator
\eq{
\label{eq:nercome}
\hat{\mx{{Z}}} \equiv \mx{U}_1\, {\rm diag} \br{\mx{U}_1^\tau\, \hat{\mx{S}}_2\, \mx{U}_1}\, \mx{U}_1^\tau\;
}
to a given subdivision of $\mx{X}$; 2. average over different compositions of $(\mx{X}_1,\mx{X}_2)$ for a given location $s$ of the split, of which there are ${ N_S \choose s}$; 3. find the optimal location of the data vector split by minimising
\eq{
\label{eq:nercome_min}
Q(s) = \left| \left| \overline{\hat{\mx{Z}}}(s) - \overline{\hat{\mx{S}}}_2(s) \right| \right|_{\rm F}^2\;,
}
where the bar denotes the average of step (2.), and where $||\mx{A}||_{\rm F}^2 = {\rm Tr} (\mx{A}\, \mx{A}^\tau)$ is the Frobenius matrix norm. An estimate for the inverse covariance is then simply provided by the inverse of the covariance estimator.

\begin{figure}
\includegraphics[height=0.95\columnwidth,angle=270,trim={0 0 2.4cm 0},clip]{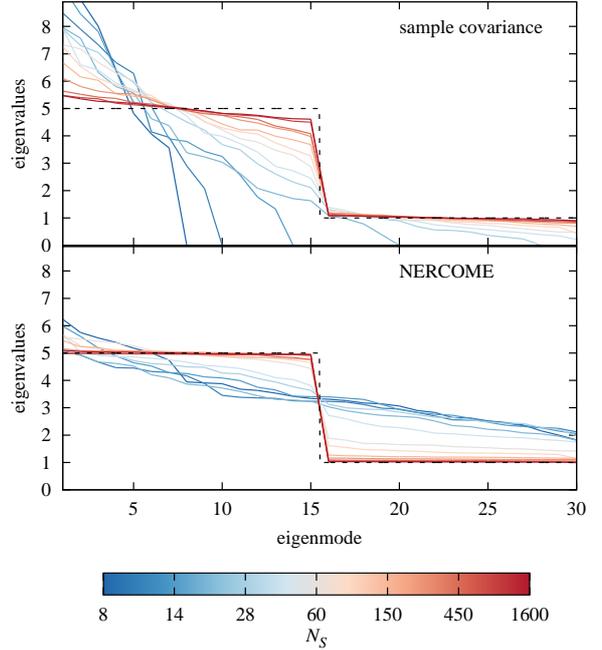}
\caption{Spectrum of eigenvalues of a $N_D=30$-dimensional covariance estimate for the toy example of uncorrelated data with standard errors of 1 and 5 for $50\,\%$ of the dataset, respectively (see black dashed line). Coloured curves result for different numbers of realisations of the data, $N_S$. \textit{Top panel}: sample covariance estimates. \textit{Bottom panel}: NERCOME estimates.}
\label{fig:eigen_toy}
\end{figure}

Equation (\ref{eq:nercome}) takes advantage of the fact that $\hat{\mx{{S}}}_1$ and $\hat{\mx{{S}}}_2$ are estimated independently from the data, so that their combination can be expected to be less adversely affected by noise when estimating the diagonal elements of the covariance. \citet{lam16} showed that $\mx{U}_1^\tau\, \hat{\mx{S}}_2\, \mx{U}_1$ is the expression for the diagonal elements that minimises the difference to the true covariance in the Frobenius norm. \citet{abadir14} demonstrated that the subsequent averaging over a moderate number of compositions of $(\mx{X}_1,\mx{X}_2)$ suppresses noise in $\hat{\mx{{Z}}}$. Here, that number is chosen to be $N_{\rm av}={\rm min}\bc{{ N_S \choose s};500}$, where in the latter case combinations are drawn at random once ${ N_S \choose s} > 3 N_{\rm av}$. Equation (\ref{eq:nercome_min}) is minimised by evaluating $Q$ at 20 equidistant steps in $s$ in the rage $\bb{0.1 N_S;0.9 N_S}$. Since $Q$ is itself a rather noisy quantity primarily through $\hat{\mx{S}}_2$, which serves as an unbiased estimate of the true covariance matrix, results for a fixed split at $s/N_S=2/3$ (meaning two thirds of the data are used to estimate $\mx{U}$) are also reported.

\begin{figure*}
\includegraphics[height=2.\columnwidth,angle=270]{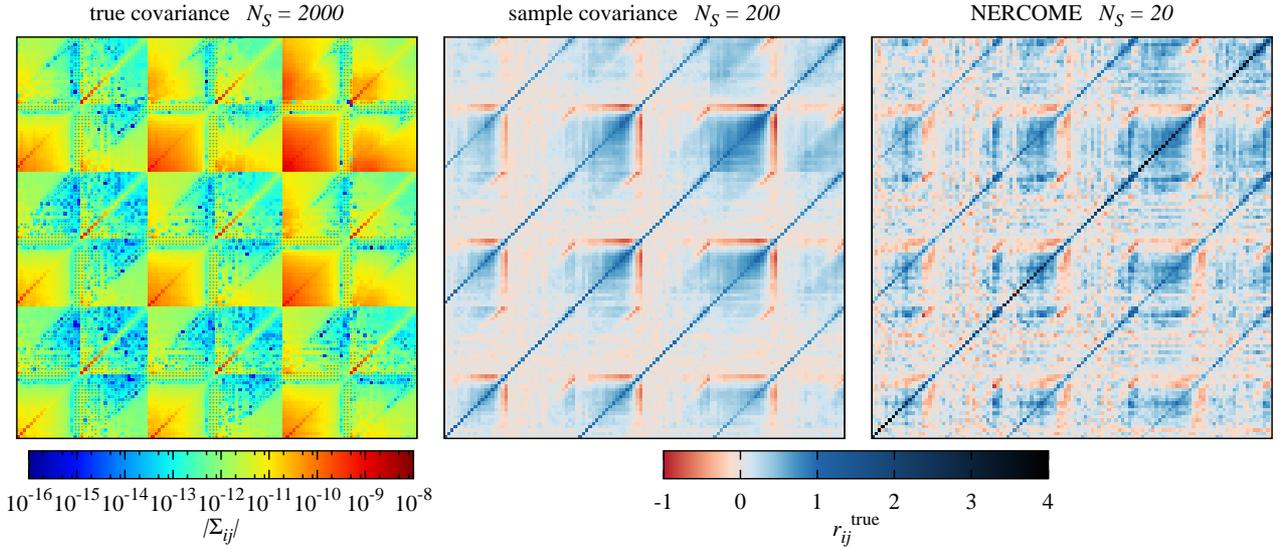}
\caption{\textit{Left panel}: \lq true\rq\ covariance (sample covariance determined from $N_S=2000$ realisations). Shown are the absolute values of the covariance elements, with negative elements indicated by the additional black markers. The block structure reflects the use of three tomographic redshift bin combinations, with $\xi_+$ and $\xi_-$ calculated for each combination. \textit{Centre panel}: sample covariance for $N_S=200$, normalised by the diagonal elements of the true covariance (see Eq. \ref{eq:rtrue}). \textit{Right panel}: Same as centre panel but for the NERCOME estimate with $N_S=20$.}
\label{fig:matrices}
\end{figure*}

NERCOME is close to ideal when the true covariance is a multiple of the identity, $a\, \mx{I}$, as asymptotically the value of $\mx{U}$ becomes irrelevant and most constraining power can be focussed on estimating the single number $a$ \citep{lam16}. It is therefore advisable to whiten the covariance by an informed choice of the normalisation, $\vek{n}$, if possible, in case an optimal estimate of $\mxg{\Sigma}$ in a mean square error sense is the goal. However, in physical applications one is usually more interested in controlling the uncertainty and bias of weighted sums of inverse covariance entries that enter likelihood analyses and weighted least squares fits. This is assessed here by using the scalar quantity $F \equiv (S/N)^2 = \vek{m}^\tau\, \mxg{\Sigma}^{-1} \vek{m}$ as the figure of merit for covariance estimation, where $\vek{m}$ is the best-guess model of the data vector. Setting $\vek{n}=\vek{m}$ for the remainder of this work, the NERCOME formalism now optimises the performance with respect to the same combination of covariance elements as those in $F$.

An incorrect choice of the model $\vek{m}$ will not per se lead to biased inference on cosmological parameters but to a potentially suboptimal performance of the NERCOME algorithm (which could then imply biases on the posterior distribution). As long as signals can be predicted to within a few tens of per cent accuracy, the uncertainty should only marginally affect the NERCOME estimate.

\section{A toy example}

The performance of NERCOME is first illustrated with a toy example, based on uncorrelated, Gaussian distributed data with standard error 5 for the first half of the data vector, and 1 for the second. The resulting spectrum of eigenvalues is shown in Fig. \ref{fig:eigen_toy} for the standard sample covariance and NERCOME. Since the matrix of eigenvectors is orthogonal and thus always well-conditioned, all ill-conditioning due to noise shows up in the eigenvalues. This is apparent for the sample covariance with small eigenvalues decreasing and large ones strongly increasing as the number of realisations decreases. Once $N_S < N_D + 2$, at least one eigenvalue vanishes so that the covariance becomes singular \citep[e.g.][]{taylor14}.

NERCOME shrinks both excessively large and small eigenvalues back towards the true values, avoiding singular values altogether. It can be shown \citep{lam16} that NERCOME estimates are positive definite with probability one (i.e. the exceptions constitute a set of measure zero) and consistent (approaching $\mxg{\Sigma}$ for $N_S \rightarrow \infty$). The shrinkage is non-linear in that different eigenvalues are shrunk by different amounts (see \citealp{pope08} for an illustration of linear shrinkage). NERCOME consistently overestimates the smallest eigenvalues for low values of $N_S$, a feature that is also present in the following more realistic example.

\section{Simulation setup}

A realistic and challenging performance test is provided by the covariance of the two-point correlation functions $\xi_\pm$ of cosmic weak lensing, measured deeply into the non-linear regime of structure formation (see \citealp{kilbinger15} for a recent review). A large suite of simulated weak lensing shear catalogues is created by producing coupled lognormal random fields from angular power spectra calculated for a vanilla flat $\Lambda$CDM cosmology and assuming a minimum lensing convergence value of $\kappa_0=-0.012$ \citep{hilbert11}. The redshift distribution of source galaxies is set to have a median of 0.8, and is split at the median into two tomographic bins. The mock survey is assumed to have an area of $25\,{\rm deg}^2$ and a source galaxy number density of $10\,{\rm arcmin}^{-2}$ per tomographic bin, with a total galaxy ellipticity dispersion of $0.35$. While the choice of survey area only leads to a rescaling of the covariance (note that the simulations have periodic boundary conditions and no masks), the number density determines the level of shot noise that contributes to the diagonal (and some sub-diagonals) of the covariance. Current and future surveys will choose their redshift binning such that at most a few galaxies per ${\rm arcmin}^2$ will be in each bin, so that the choice above will lead to larger cross-correlations than expected in real-world applications.

From the 2000 mock realisations created in total, the shear correlation functions $\xi_\pm$ are measured for all three redshift bin combinations in 20 angular bins, logarithmically spaced between 1 and 180$\,$arcmin, with the tree code ATHENA \citep{kilbinger_athena}, constituting a total data vector of length $N_D=120$. The resulting covariance is shown in Fig. \ref{fig:matrices} and is challenging for non-standard estimators in that it has a high condition number ($\sim 3000$), a high level of correlation (numerous off-diagonal elements with correlation close to $\pm 1$; see the centre panel of Fig. \ref{fig:matrices} ), and a complex structure including several discontinuities.

\section{Performance}

\begin{figure}
\includegraphics[height=0.95\columnwidth,angle=270,trim={0 0 2.4cm 0},clip]{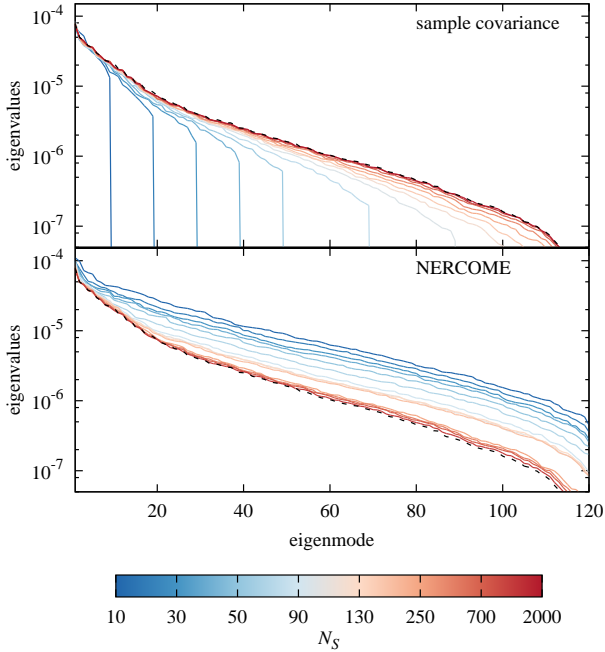}
\caption{Same as Fig. \ref{fig:eigen_toy}, but for the cosmic weak lensing covariance shown in Fig. \ref{fig:matrices}. The \lq true\rq\ spectrum as measured from the sample covariance with $N_S=2000$ realisations is given by the black dashed line. Note the logarithmic scaling of the ordinate axes.}
\label{fig:eigen_sim}
\end{figure}

The eigenspectra for the simulated covariance, shown in Fig. \ref{fig:eigen_sim}, are qualitatively similar to the toy case; NERCOME estimates are consistent, remain positive definite, and display a positive bias across the spectrum. This is reflected in an over-estimation of covariance elements, particularly the diagonal ones, that increases as $N_S$ drops. 

This is illustrated in the right panel of Fig. \ref{fig:matrices}, where a correlation matrix normalised with respect to the diagonal elements of the \lq true\rq\ covariance (sample covariance for $N_S=2000$),
\eq{
\label{eq:rtrue}
r_{ij}^{\rm true} \equiv C_{ij}/\sqrt{C_{ii}^{\rm true}\,C_{jj}^{\rm true} }\;,
}
is shown. For $N_S=20$, diagonal elements are larger by a factor of up to 4. Otherwise, all main features in the correlation matrix have been reproduced, despite the small number of realisations (cf. the sample covariance in the centre panel). This trend persists irrespectively of whether $\vek{n}$ is set to unity or to $\vek{m}$, hence NERCOME is a poor choice of estimator if the covariance itself is the desired outcome of the estimation process.

\begin{figure}
\includegraphics[height=0.95\columnwidth,angle=270,trim={0 0 3.5cm 0},clip]{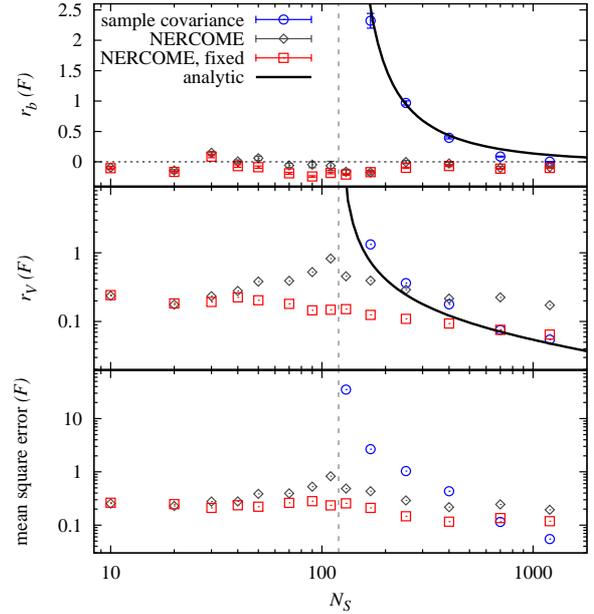}
\caption{Relative bias (\textit{top panel}), variance (\textit{centre panel}), and mean square error (\textit{bottom panel}) for the squared signal-to-noise, $F$, as a function of the number of data realisations, $N_S$, used to estimate the covariance. Blue circles (grey diamonds, red squares) correspond to using the sample covariance (default NERCOME, NERCOME with a split fixed at $s/N_S=2/3$). The black solid curves show the analytic expectation for the sample covariance. The vertical grey dashed line indicates $N_S=N_D=120$.}
\label{fig:sn}
\end{figure}

In Fig. \ref{fig:sn} the performance in terms of the signal-to-noise, $F$, is shown. Under the assumption that the sample covariance is Wishart distributed, the distributions of its inverse and linear mappings thereof can be calculated analytically. It follows that an estimate of $F$ derived from using the inverse of the sample covariance is distributed according to $\hat{F} \sim {\rm Inv-}\!\chi^2(F,N_S-N_D)$ \citep{eaton07}. This allows for the calculation of the relative bias and rms noise of $F$, given by
\eqa{
r_b &\!\!\!\!\!\equiv\!\!\!\!\!& \frac{\ba{\hat{F}}}{F} - 1 = \frac{N_S-1}{N_S-N_D-2}\;; \\
r_V &\!\!\!\!\!\equiv\!\!\!\!\!& \frac{\sqrt{\ba{\br{\hat{F}-\ba{\hat{F}}}^2}}}{F} = \frac{\sqrt{2}\; (N_S-1)}{\sqrt{N_S-N_D-4}\; (N_S-N_D-2)}\,;
}
see \citet{taylor14} for the details regarding moment calculations. For a given $N_S$, the mean and variance of $\hat{F}$ entering $r_b$ and $r_V$ are calculated via a delete-one jackknife. The analytic predictions agree excellently with the simulation results, validating the approach. Both bias and variance diverge as $N_S \rightarrow N_D$, with $\hat{\mx{S}}$ moving ever closer to becoming singular. The standard NERCOME estimator performs very well, displaying small, marginally significant bias over the range of $N_S$ probed and a relative rms error that peaks around $N_S \approx N_D$ and reduces to $\sim 0.2$ for both large and small $N_S$. For $N_S > 400$, the variance of $\hat{F}$ via NERCOME surpasses that using $\hat{\mx{S}}$, which could be beaten down by increasing $N_{\rm av}$, but the sample covariance estimator is the more efficient choice in this regime anyway. The location of the split fluctuates typically between $s/N_S=0.5$ and $1$, as $N_S$ varies, and tends to larger values in the regime where $r_V$ peaks. An alternative run of NERCOME with $s/N_S$ fixed at $2/3$, also shown in Fig. \ref{fig:sn}, returns slightly smaller and almost constant mean square error, at the price of a somewhat larger bias contribution for some values of $N_S$. Overall, NERCOME down to $N_S=10$ (cf. Fig. \ref{fig:matrices}, right panel) is competitive with the sample covariance estimator at $N_S \sim 500$ in terms of the mean square error.

\begin{figure}
\includegraphics[height=0.9\columnwidth,angle=270,trim={0 0 0 0},clip]{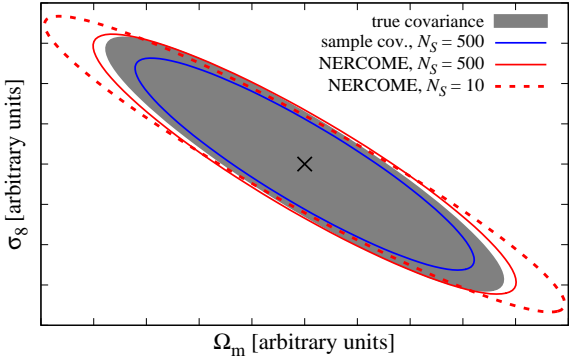}
\caption{Comparison of parameter constraints in the $\Omega_{\rm m} ? \sigma_8$ plane, using different covariance estimates in a Fisher matrix calculation, as indicated in the legend.}
\label{fig:fisher}
\end{figure}

To assess the impact on cosmological posteriors, the different covariance estimates are inserted into a Fisher matrix forecast (following \citealp{tegmark97}) for the two most strongly constrained parameters for the toy weak lensing survey outlined above, $\Omega_{\rm m}$ and $\sigma_8$, with results shown in Fig. \ref{fig:fisher}. For $N_S = 500$, NERCOME is very close to the result for the \mbox{\lq true\rq} covariance, while the sample covariance underestimates the width of the posterior, corresponding to an overestimate of $F$. The latter bias can largely be removed by applying the analytic correction of the mean bias as proposed by \citet{hartlap06}; cf. Fig. \ref{fig:sn}, top. The $N_S = 10$ NERCOME estimate accurately reproduces the degeneracy direction and the size of the minor axis of the posterior ellipse, but biases the major axis moderately high. Note that the choice of normalisation $\vek{n}$ in this work is optimal for $F$ but not necessarily so for elements of the Fisher matrix.

\section{Conclusions}

This work marks the first application of a non-linear shrinkage estimator of covariance in an astrophysical context, using the realistic example of a tomographic cosmic weak lensing analysis that features high condition number, high levels of correlation, and a complex structure of the covariance matrix. After rescaling with a model of the data vector, the NERCOME estimator is able to estimate a function of the inverse covariance that has the same form as a Gaussian log-likelihood with subdominant bias and with variance that scales only mildly with the number of realisations of the data vector. Well-conditioned covariance estimates in the regime of much fewer realisations than data points are readily achieved. Compared to the standard sample covariance estimator, a factor 50 less realisations are required to achieve the same mean square error, without any assumptions on the statistical properties of the data or the form of the covariance matrix, beyond those made for the sample covariance estimator (such as independently and identically distributed data).

The NERCOME estimator is consistent and almost surely positive definite. Its algorithm is simple, mainly consisting of eigenvalue decompositions, and trivially parallelisable in the subsequent averaging and split optimisation steps. On a single core of a standard Unix work station NERCOME takes $\sim 6\,$s wall-clock time per split location for the $120$-dimensional data vector and setup considered in this work, so adds negligible runtime to an analysis pipeline. A downside is the lack of a priori control over the bias of the estimator, which depends on the details of the structure of the covariance matrix, an issue which is shared with most alternative estimators of covariance \citep[e.g.][]{pope08,paz15}. However, fast approximate simulations based on random fields \citep[see e.g.][]{xavier16} are well suited to assess these biases and optimise the free parameters of the estimators. Further work is required to assess the impact of these alternative covariance estimators on the fidelity of posteriors. A first cursory test with NERCOME revealed moderate bias for very low $N_S$ along the least constrained direction in parameter space.

Further suppression of adverse noise effects from covariance estimation can be achieved by explicitly incorporating prior information, which can range from the assertion of smoothness in the elements of (sub-)matrices to physically motivated effective models of the full covariance with a small level of residual degrees of freedom (see \citealp{mandelbaum13,oconnell15,pearson16} for recent applications). A combination of such covariance modelling with the principles of non-linear shrinkage estimation as employed in NERCOME is a promising avenue. Likewise, a combination of shrinkage with resampling techniques directly applied to the data could potentially obviate the need for simulated data altogether, provided the challenges of bias induced by long-range spatial correlations can be overcome \citep{norberg09,friedrich16}. Since a NERCOME-like estimator does not set any requirements on the structure of the data vector (such as smoothness or a notion of distance), it can also readily be combined with compression of the data vector as a pre-processing step (see \citealp{taylor13} and references therein, as well as \citealp{zablocki16} for a recent example).

An implementation of NERCOME in C is made available with this publication.\footnote{http://www.star.ucl.ac.uk/$\sim$joachimi/publications.html}

\section*{Acknowledgements}

The author thanks C. Lam for helpful discussions and the anonymous referee for a constructive report. BJ acknowledges support by an STFC Ernest Rutherford Fellowship, grant reference ST/J004421/1.




\bibliographystyle{mnras}


\bsp	
\label{lastpage}
\end{document}